\documentclass[aps,prl,superscriptaddress,amsmath,amssymb,preprintnumbers,showpacs,twocolumn,reprint,dvipdfmx]{revtex4-1}

\usepackage{amssymb}
\usepackage{graphicx}
\usepackage{dcolumn}
\usepackage{bm}
\usepackage{color}
\usepackage{subfig}

\begin{document}

\newcommand{\beq}{\begin{equation}}
\newcommand{\eeq}{\end{equation}}
\newcommand{\barr}{\begin{eqnarray}}
\newcommand{\earr}{\end{eqnarray}}

\def\bra#1{\langle{#1}|}
\def\ket#1{|{#1}\rangle}
\def\sinc{\mathop{\text{sinc}}\nolimits}
\def\cV{\mathcal{V}}
\def\cH{\mathcal{H}}
\def\cT{\mathcal{T}}
\renewcommand{\Re}{\mathop{\text{Re}}\nolimits}
\newcommand{\tr}{\mathop{\text{Tr}}\nolimits}

\definecolor{dgreen}{rgb}{0,0.5,0}
\newcommand{\green}{\color{dgreen}}
\newcommand{\BLUE}[1]{\text{\color{blue} #1}}
\newcommand{\GREEN}[1]{\textbf{\color{green}#1}}
\newcommand{\REV}[1]{\textbf{\color{red}[[#1]]}}
\newcommand{\KY}[1]{\textbf{\color{dgreen}[[#1]]}}

\def\HN#1{{\color{magenta}#1}}
\def\DEL#1{{\color{yellow}#1}}

\title{Analytic estimation of transition between instantaneous eigenstates of quantum two-level system}


\author{Takayuki Suzuki}
\author{Hiromichi Nakazato}
\email{hiromici@waseda.jp}
\affiliation{Department of Physics, Waseda University, Tokyo 169-8555, Japan}

\author{Roberto Grimaudo}
\affiliation{Dipartimento di Fisica e Chimica dell'Universit\`a di Palermo, Via Archirafi, 36, I-90123 Palermo, Italy}
\affiliation{INFN, Sezione di Catania, Catania, Italy}
\author{Antonino Messina}%
\affiliation{Dipartimento di Matematica ed Informatica dell'Universita di Palermo, Via Archirafi 34, I-90123 Palermo, Italy}
\affiliation{INFN, Sezione di Catania, Catania, Italy}
\begin{abstract}
Transition amplitudes between instantaneous eigenstates of quantum two-level system are evaluated analytically on the basis of a new parametrization of its evolution operator, which has recently been proposed to construct exact solutions.
In particular, they are estimated when the Hamiltonian varies infinitesimally slowly.  
The results, not only confirm the adiabatic theorem in the adiabatic limit, but also bring us with an analytic estimation of the adiabatic approximation.
The condition under which no transition between different instantaneous eigenstates is allowed is also clarified. 
\end{abstract}

\maketitle


A quantum system described by a time-dependent nondegenerate Hamiltonian $H(t)$, such that $[H(t),H(t')]\neq 0$, does not possess stationary states, since, in general, the mean value of an observable changes in time. 
The eigenstates of $H(t)$ (instantaneous eigenstates) depend on time and evolve into those states that are no longer the corresponding eigenstates of the Hamiltonian at that time instant.
The deviations from the eigenstates, however, become  negligible under the situation where the so-called adiabatic theorem \cite{BornFock1928,Kato1950,Messiah1962} is applicable. 
The theorem states that the $n$th instantaneous eigenstate evolves remaining with continuity in the $n$th eigenstate at any time instant. 
The condition for such an occurrence in a somewhat intuitive sense is that the quantum dynamics of the system is governed by a  Hamiltonian that changes vanishingly slowly in time.

The adiabatic theorem is so important to have been constituting a basis of  wide  research fields \cite{LZ,GL,Berry,QC,QZ}.
In its proofs, the deviations from the adiabatic limit, i.e., an infinitesimal rate of variation over an infinite time interval $T\to\infty$ yielding a finite nonvanishing variation of Hamiltonian, are estimated to be suppressed by $1/T$.
The adiabatic theorem holds irrespectively of details of the system under consideration, thus making it applicable to a wide class of quantum systems.
On the other hand, any physically realizable process can not be in the adiabatic limit because the time duration $T$ can not be made infinite and the physical process can be considered at most approximately adiabatic.
It is, therefore, of practical relevance to understand how well it is approximated as adiabatic \cite{MarzlinSanders2004,Tong2007,Du2008,WangPlenio2016} and to estimate what the possible deviations from the adiabatic limit are.
The theorem gives little information on these issues and a precise estimation of such deviations would require an exact knowledge of the dynamics.
It is thus desirable to have an access to exact solutions of the dynamics and then natural to expect to have new insights on these issues once a new strategy to construct analytical solutions has been proposed.
In this respect, it is worth while stressing that such a strategy for obtaining exact solutions for quantum two-level systems was proposed recently \cite{BarnesDasSarma,MessinaNakazato2014} and actually several exactly solvable examples, exploitable also for interacting qudits, have newly been obtained \cite{Grimaudo2016} according to the strategy \cite{MessinaNakazato2014}.

The purpose of this Letter is threefold.
First, the transition amplitudes between instantaneous eigenstates of the general time-dependent Hamiltonian for quantum two-level system are shown to be expressed analytically in terms of appropriate quantities that parametrize the dynamics.
It is remarkable that such analytical expressions are available even if the dependence on the parameters in the Hamiltonian is implicit.
Second, the transition amplitude from one of the instantaneous eigenstates to the other one is shown to vanish in the adiabatic limit (or the infinite time-interval $T\to\infty$ limit), confirming the adiabatic theorem and its deviation from the adiabatic limit (i.e., transition for a large but finite time interval $T<\infty$) is evaluated analytically.
An explicit solvable example that can represent an adiabatic process is supplied to illustrate the general characteristics.
Finally, the condition under which such a transition is suppressed is examined and discussed.

Let a quantum two-level system be described by a time-dependent Hamiltonian
\begin{equation}
H(t)=\begin{pmatrix}\Omega(t)&\omega(t)\\
                                    \omega^*(t)&-\Omega(t)
          \end{pmatrix},
\quad\omega(t)=|\omega(t)|e^{i\phi_\omega(t)},
\end{equation}
where $\Omega$ and $\omega$ are time-dependent real and complex functions, respectively.
The instantaneous eigenvalues $E_\pm(t)=\pm\sqrt{\Omega^2(t)+|\omega(t)|^2}$ and eingenstates $|\pm\rangle_t$,
\begin{equation}
|+\rangle_t=\begin{pmatrix}e^{{i\over2}\phi_\omega}\cos{\theta\over2}\\e^{-{i\over2}\phi_\omega}\sin{\theta\over2}\end{pmatrix},\quad
|-\rangle_t=\begin{pmatrix}e^{{i\over2}\phi_\omega}\sin{\theta\over2}\\-e^{-{i\over2}\phi_\omega}\cos{\theta\over2}\end{pmatrix},
\end{equation}
are both time dependent,  where $\tan\theta=|\omega|/\Omega$ \cite{t-dependence}.
According to the strategy proposed in \cite{MessinaNakazato2014}, the evolution operator $U(t)$ 
is given by
\begin{align}
U&=\begin{pmatrix}a&b\\-b^*&a^*\end{pmatrix},\nonumber\\
a&=\cos\chi e^{-{i\over2}(\Theta-\phi_\omega+\phi)},\,
b=-i\sin\chi e^{-{i\over2}(\Theta-\phi_\omega-\phi)},
\label{eq:a&b}
\end{align}
where $\Theta$ is an arbitrary function of time with $\Theta(0)=\phi_\omega(0)-\phi(0)=0$ and
\begin{align}
\chi&=\int_0^tdt'{|\omega(t')|\over\hbar}\cos\Theta(t'),\label{eq:chi}\\
\phi&=\int_0^tdt'{2|\omega(t')|\over\hbar}{\sin\Theta(t')\over\sin2\chi(t')}+\phi_\omega(0).
\label{eq:phi}
\end{align}
The function $\Omega$, representing the longitudinal magnetic field in the case of spin $1/2$, is specified by
\begin{equation}
\Omega={\hbar\over2}(\dot\Theta-\dot\phi_\omega)+|\omega|\sin\Theta\cot2\chi.
\label{eq:Omega}
\end{equation}
The original idea of \cite{MessinaNakazato2014} is such that, if $\Omega$ is so adjusted to satisfy (\ref{eq:Omega}) for an arbitrarily given $\omega$ and a parameter function $\Theta$, then $U$ is explicitly given in terms of them in (\ref{eq:a&b}).
One may regard (\ref{eq:a&b}) as a new parametrization of $U$.
In fact, if the evolution operator $U$ is so parametrized by $\chi,\Theta$ and $\phi$ as in (\ref{eq:a&b}),  the equations (\ref{eq:chi})-(\ref{eq:Omega}) are enough to make $U$ satisfy the Schr\"odinger equation.

It is convenient for the later use to rewrite the evolution operator $U$ in a compact form
\begin{align}
U&=\begin{pmatrix}\cos\chi e^{-{i\over2}(\Theta+\phi-\phi_\omega)}&-i\sin\chi e^{-{i\over2}(\Theta-\phi-\phi_\omega)}\\
                               -i\sin\chi e^{{i\over2}(\Theta-\phi-\phi_\omega)}&\cos\chi e^{{i\over2}(\Theta+\phi-\phi_\omega)}\end{pmatrix}\nonumber\\
&=e^{{i\over2}\sigma_z\phi_\omega}e^{-i\Phi\bm n\cdot\bm\sigma}e^{-{i\over2}\sigma_z\phi_\omega(0)},
\end{align}
where $\bm\sigma=(\sigma_x,\sigma_y,\sigma_z)$ is the Pauli matrix vector.
We have introduced a unit vector $\bm n=(n_x,n_y,n_z)$  
with $n_x=\sin\xi\cos\varphi_-={\sin\chi\over\sin\Phi}\cos\varphi_-$,
$n_y=\sin\xi\sin\varphi_-={\sin\chi\over\sin\Phi}\sin\varphi_-$ and
$n_z=\cos\xi={\cos\chi\over\sin\Phi}\sin\varphi_+$, and $\cos\Phi=\cos\chi\cos\varphi_+$, where angles are defined as $\varphi_\pm={1\over2}(\Theta\pm\bar\phi),\,\bar\phi\equiv\phi-\phi_\omega(0)$.
Then it is straightforward to calculate transition amplitudes between instantaneous eigenstates.
For example, the transition from the initial state $|+\rangle_0$ to the other eigenstate $|-\rangle_t$ occurs with an amplitude ($\theta_0=\theta(0)$)
\begin{align}
&{}_t\langle{-}|U(t)|+\rangle_0\nonumber\\
&=\cos\chi\cos\varphi_+\sin{\theta-\theta_0\over2}-\sin\chi\sin\varphi_-\cos{\theta-\theta_0\over2}\nonumber\\
&\;+i\Bigl(\sin\chi\cos\varphi_-\cos{\theta+\theta_0\over2}-\cos\chi\sin\varphi_+\sin{\theta+\theta_0\over2}\Bigr).
\label{eq:A-+}
\end{align}
Similarly, the other amplitudes are concisely expressed in terms of trigonometric functions of $\chi$, $\varphi_\pm$ and $\theta$ \cite{Suzuki2018}.
Notice that these expressions are exact and no approximation is involved.

Observe that the amplitudes are given by quantities at time $t$ and $0$ and are apparently not dependent on time-derivatives, while the adiabatic theorem dictates that the transition amplitude (\ref{eq:A-+}) vanishes in the adiabatic limit.
This apparently puzzling observation may be resolved by considering another parametrization.
Introduce new parameters $x$ and $y$ by
\begin{equation}
x\equiv\tan\chi,\quad y={|\omega|\over\hbar}\sin\Theta,\quad x(0)=y(0)=0.
\end{equation}
Then the quantities in the Hamiltonian are expressed as
\begin{align}
|\omega|&=\hbar\sqrt{y^2+\Bigl({\dot x\over1+x^2}\Bigr)^2},\\
\Omega&={\hbar\over2}(\dot\Theta-\dot\phi_\omega)+{\hbar y\over2}\Bigl({1\over x}-x\Bigr),
\end{align}
while 
\begin{equation}
\tan\Theta={1+x^2\over\dot x}y,\quad
\dot\phi=y\Bigl({1\over x}+x\Bigr).
\label{eq:tanTheta,dotphi}
\end{equation}
Now consider the situation where the Hamiltonian varies very slowly in time.
Such a situation can be realized when parameters $x$ and $y$ are both slowly varying functions of time.
Even though this is not the only possible case to realize slowly varying Hamiltonians, for the sake of simplicity and clarity we confine ourselves to such cases.
Since values of $x$ and $y$ need not necessarily be small even though they vanish at $t=0$, the parameter $\Theta$ is almost always quite close to $\pi/2$ except for the initial short period of time starting from $\Theta(0)=0$.
Set $\Theta=\pi/2-\epsilon$, then 
\begin{equation}
\epsilon=\tan^{-1}\Bigl({1\over y}{\dot x\over1+x^2}\Bigr)
\end{equation}
is an infinitesimal quantity $\epsilon\propto\dot x\sim0$ for all times except for the initial transient times where $\epsilon\sim\pi/2$ and the quantities appearing in (\ref{eq:A-+}) are re-expressed in terms of $\epsilon$.
These quantities are approximated, for small $\epsilon$ and $\dot\phi_\omega$, as
\begin{align}
\cos\varphi_\pm&\sim\cos\bigl({\pi\over4}\pm{\bar\phi\over2}\bigr)+{\epsilon\over2}\sin\bigl({\pi\over4}\pm{\bar\phi\over2}\bigr),\\
\sin\varphi_\pm&\sim\sin\bigl({\pi\over4}\pm{\bar\phi\over2}\bigr)-{\epsilon\over2}\cos\bigl({\pi\over4}\pm{\bar\phi\over2}\bigr),\\
\tan\theta&\sim\tan2\chi\bigl(1+{\tan2\chi\over2y}(\dot\epsilon+\dot\phi_\omega)\bigr),
\end{align}
while
\begin{equation}
\tan\theta_0=-{\dot x(0)\over\dot\epsilon(0)+{1\over2}\dot\phi_\omega(0)}\sim-{\dot x(0)\over\dot\epsilon(0)}
\end{equation}
is also considered small if $\dot x(0)$ is infinitesimal and $\dot\epsilon(0)\sim-{\dot y(0)\over\dot x(0)}$ is a finite, non-vanishing quantity (recall $\epsilon(t)\sim0$ for $t>0$ while $\epsilon(0)=\pi/2$).

It is evident that at the lowest order, that is, in the adiabatic limit, the transition amplitude (\ref{eq:A-+}) vanishes
\begin{equation}
{}_t\langle-|U(t)|+\rangle_0\sim0,\quad\forall t\ge0,
\end{equation}
since we can set
\begin{equation}
\cos\varphi_\pm\sim\sin\varphi_\mp,\quad
{\theta\pm\theta_0\over2}\sim\chi.
\end{equation}
This is the manifestation of the adiabatic theorem in the situation here considered.
Needless to say, since the variation rates are infinitesimal, to achieve a finite dynamical change, we need an infinite time interval $T\to\infty$.
In other words, the variations rates are proportional to $1/T$, i.e., $\dot x,\dot y,\dot\phi_\omega\propto1/T\to0$.
Remark that this confirmation of the theorem is based on the explicit knowledge of the dynamics, which is made possible by the use of its new parametrization.
It is also stressed that the result is not restricted to a particular solvable case, but covers  a wide range of cases where the functions in the Hamiltonian $\Omega$ and $\omega$ are characterized by infinitesimally slowly varying functions of time $x$ and $y$.
Of course, there are cases where the present characterization of slowly varying $\Omega$ and $\omega$ is not applicable
, for which still other parametrization would be necessary.

Deviations from the adiabatic limit are reflected in the nonvanishing transition amplitude (\ref{eq:A-+}) when the Hamiltonian changes with finite rates, or whenever we consider a finite dynamical change within a finite interval of time $T<\infty$.
Apparently in this case the transition amplitude (\ref{eq:A-+}) is proportional to $\epsilon\propto\dot x,\dot\phi_\omega$ and $\dot y$.
We first approximately evaluate trigonometric functions of $(\theta\mp\theta_0)/2$ as
\begin{align}
\cos{\theta\mp\theta_0\over2}
&\sim\cos\chi-{1\over2}\Bigl(\mp{\dot x^2(0)\over\dot y(0)}+{\sin^22\chi\over2y}\dot\phi_\omega\Bigl)\sin\chi,\\
\sin{\theta\mp\theta_0\over2}
&\sim\sin\chi+{1\over2}\Bigl(\mp{\dot x^2(0)\over\dot y(0)}+{\sin^22\chi\over2y}\dot\phi_\omega\Bigl)\cos\chi,
\end{align}
where for simplicity we have assumed that $\dot\epsilon$ is of higher order in the variation and neglected it.
Expressing the trigonometric functions of $\chi$ as functions of $x=\tan\chi$, we obtain 
\begin{align}
{}_t\langle-|U(t)|+\rangle_0
&\sim{x\dot x\over y(1+x^2)^2}e^{i({\pi\over4}-{\bar\phi\over2})}
-{\dot x^2(0)\over2\dot y(0)}e^{i({\pi\over4}+{\bar\phi\over2})}\nonumber\\
&\quad+{x^2\over y(1+x^2)^2}\dot\phi_\omega e^{-i({\pi\over4}+{\bar\phi\over2})}.
\end{align}

The transition probability from the eigenstate $|+\rangle_0$ at time $t=0$ to the other one at a later time $|-\rangle_t,\,t>0$ is vanishingly small when the variation rate of the Hamiltonian is infinitesimal and reads as
\begin{align}
&\bigl|_t\langle-|U(t)|+\rangle_0\bigr|^2\nonumber\\
&\sim \left({x\dot x\over y(1+x^2)^2}\right)^2+\left({\dot x^2(0)\over2\dot y(0)}\right)^2+\left({x^2\over y(1+x^2)^2}\dot\phi_\omega\right)^2\nonumber\\
&\quad+{\dot x^2(0)\over\dot y(0)}{x\over y(1+x^2)^2}\bigl(x\dot\phi_\omega\sin\bar\phi-\dot x\cos\bar\phi\bigr).
\end{align}
Remark again that the quantities in the Hamiltonian $\Omega$ and $|\omega|$ are expressed as functions of $x,y$ and $\phi_\omega$ and that this expression is valid when variation rates of the latters are infinitesimal.

As an example, we may consider a particular case of physical interest.
Choose the functions $x$ and $y$ as
\begin{equation}
x=\sin\alpha t,\quad y=\nu_0\sin\alpha t
\end{equation}
with parameters $\nu_0>0$ and $\alpha\propto1/T$.
Then we have
\begin{align}
|\omega|&=\hbar\sqrt{\nu_0^2\sin^2\alpha t+\Bigl({\alpha\cos\alpha t\over1+\sin^2\alpha t}\Bigr)^2},\\
\Omega&={\hbar\over2}(\dot\Theta-\dot\phi_\omega)+{\hbar\nu_0\over2}\cos^2\alpha t,
\end{align}
where $\tan\Theta={\nu_0\over\alpha}\bigl(1+\sin^2\alpha t\bigr)\tan\alpha t$.
Observe that the physical system under consideration could be well approximated for small $\alpha,\dot\phi_\omega\sim0$ by   
\begin{equation}
|\omega|\sim\hbar\nu_0|\sin\alpha t|,\quad\Omega\sim{\hbar\nu_0\over2}\cos^2\alpha t
\label{eq:omegas}
\end{equation}
except for the initial transient times, where we have
\begin{equation}
|\omega(0)|=\hbar\alpha,\quad\Omega(0)=\hbar\nu_0.
\label{eq:initialomegas}
\end{equation}

By setting $\alpha T=\pi/2$, we are going to consider a physical process where the magnetic field (for spin 1/2 case), starting from an approximately longitudinal one (\ref{eq:initialomegas}) if $\alpha=\pi/2T\ll\nu_0$, varies slowly in time characterized by the trigonometric functions (\ref{eq:omegas}), reaching an approximately transversal one 
\begin{equation}
|\omega(T)|=\hbar\nu_0,\quad\Omega(T)={\hbar\over2}\Bigl({\alpha^2\over2\nu_0}-\dot\phi_\omega(T)\Bigr)\ll\hbar\nu_0.
\end{equation}
Clearly the adiabaticity of the process  is characterized by the inequalities
\begin{equation}
{\alpha\over\nu_0}={\pi\over2T\nu_0}\ll1,\quad {\dot\phi_\omega\over\nu_0}\ll1.
\label{eq:adiabaticity}
\end{equation} 
Observe that in this case the parameter $\nu_0$ can be viewed as a characteristic frequency relevant to the energy of the system, while $\alpha$ is the fundamental frequency governing the variation of the Hamiltonian.
Since we are given an analytical expression for every quantity (for example, $\bar\phi(t)=\nu_0(6\alpha t-\sin2\alpha t)/4\alpha$), the transition amplitudes and probabilities are calculated without any approximation.
The following figures show the behavior of transition probability $|_t\langle-|U(t)|+\rangle_0|^2$ as a function of time $t$ for several values of $\nu_0T$, together with that of longitudinal ($\Omega$) and transversal ($|\omega|$) magnetic fields. 
\begin{figure}[hbt]
\centering
\includegraphics[scale=.22]{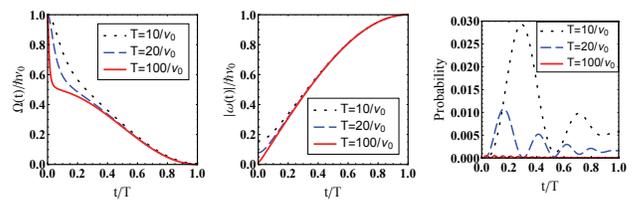}
\caption{(Color online) Magnetic fields $\Omega$ (left) and $|\omega|$ (middle) and transition probability $|{}_t\langle-|U(t)|+\rangle_0|^2$ (right) as functions of $t$ for $\nu_0T=10$ (dotted black lines), $20$ (broken blue lines) and $100$ (solid red lines).}
\end{figure}
It is evident that in the large $\nu_0T\gg1$ limit, the transition probability becomes vanishingly small.\goodbreak
 
The analytical expression of transition amplitude (\ref{eq:A-+}) enables us to investigate the condition under which no transition between different eigenstates at any time is allowed, ${}_t\langle-|U(t)|+\rangle_0=0,\,\forall t\ge0$.
The condition requires the following relations to hold
\begin{equation}
\tan{\theta+\theta_0\over2}=\tan\chi{\cos\varphi_-\over\sin\varphi_+},\;
\tan{\theta-\theta_0\over2}=\tan\chi{\sin\varphi_-\over\cos\varphi_+}.
\end{equation}
They are combined to yield 
\begin{align}
\tan\theta&={2\tan\chi\cos\bar\phi\over\sin\Theta\cos\bar\phi(1-\tan^2\chi)+\cos\Theta\sin\bar\phi(1+\tan^2\chi)}\nonumber\\
&={2\sqrt{y^2+\bigl({\dot x\over1+x^2}\bigr)^2}\over y\bigl({1-x^2\over x}\bigr)+{\dot x\over x}\tan\bar\phi},
\end{align}
from which we obtain
\begin{equation}
\dot\phi_\omega=\dot\Theta-{\dot x\over x}\tan\bar\phi,
\end{equation}
and
\begin{equation}
\tan\theta_0={2\tan\chi\over\tan\Theta\cos\bar\phi(1+\tan^2\chi)+\sin\bar\phi(1-\tan^2\chi)},
\end{equation}
which can be reduced to the following differential equation 
\begin{equation}
{d\over dt}\Bigl({x\over1+x^2}\sin\bar\phi+{1\over\tan\theta_0}{1\over1+x^2}\Bigr)=0.
\label{eq:de}
\end{equation}
The solution is easily found to be $\sin\bar\phi=x/\tan\theta_0$, which leads to $\theta=\theta_0$ and $\dot\phi_\omega=0$ after transforming the solution to the relation between functions $x$ and $y$.
The result is trivial in the sense that the instantaneous eigenstates do not at all evolve in time.

In order to allow them to evolve, we need to admit to add a finite quantity which is not constant but whose velocity can be neglected.
(This concept is in accord with the notion of limit of adiabaticity, i.e., a finite change with an infinitesimal velocity.)
Now the solution of (\ref{eq:de}) would be parametrized as
\begin{equation}
\sin\bar\phi={x\over\tan\theta_0}\Bigl(1+{1+x^2\over x^2}c\Bigr),
\label{eq:sinphibar}
\end{equation}
where the function $c$ ($c(0)=0$) is assumed to be almost constant in time but takes a finite nonvanishing value at $t>0$, like $c=c(t/T)$ for large $T$ ($0\le t\le T$).
We will neglect terms proportional to $\dot c\propto1/T$, but keep those proportional to $\dot x$ or $c$.
The above solution is converted to the relation between $x$ and $y$
\begin{equation}
y={\dot x\over1+x^2}{x\over\cos\bar\phi\tan\theta_0}\Bigl(1-{1-x^2\over x^2}c\Bigr),
\end{equation}
which yields
\begin{align}
\Omega&=\hbar{\dot x\over(1+x^2)\cos\bar\phi}{1+2c\over\tan\theta_0},\label{eq:OmegaAd}\\
\omega&=\hbar{\dot x\over(1+x^2)\cos\bar\phi}\sqrt{1-{4c(1+c)\over\tan^2\theta_0}}e^{i\phi_\omega},\;\dot\phi_\omega\sim0.
\label{eq:omegaAd}
\end{align}
The consistency condition $\tan\theta=|\omega|/\Omega$ requires that 
\begin{equation}
c={1\over2}\Bigl({\cos\theta\over\cos\theta_0}-1\Bigr).
\end{equation}
Thus we have found that the condition that no transition between different eigenstates occurs requires that the Hamiltonian has to be parametrized as above with arbitrary functions $x,\phi_\omega$ and $c$.
The condition $|\dot x|\gg|\dot c|\sim0$ is equivalent to $\dot\theta\sim0$, implying, together with $\dot\phi_\omega\sim0$, that the instantaneous eigenstates have to vary infinitesimally slowly in time, though the Hamiltonian itself can vary arbitrarily \cite{WangPlenio2016}.  
It would be interesting to see that Eq.\ (\ref{eq:OmegaAd}) can actually be integrated, thus assuring that such a parametrization is always possible, to yield 
\begin{equation}
\tan\bar\phi={\cos\theta_0\over\sin(2\zeta)}\Bigl({\tan\theta_0\over\tan\theta}-\cos(2\zeta)\Bigr),
\end{equation}
where $\zeta=\int_0^tds\sqrt{\Omega^2(s)+|\omega(s)|^2}/\hbar$.
This, together with (\ref{eq:sinphibar}), entails to specify $x$ in terms of $\Omega$ and $|\omega|$.
It is also possible to show that when the Hamiltonian is characterized by the above $\Omega$ and $\omega$, (\ref{eq:OmegaAd}) and (\ref{eq:omegaAd}), the transition amplitude (\ref{eq:A-+}) vanishes, provided $\dot c\sim0$, i.e., $\dot\theta\sim0$.
   
In summary, the adiabatic theorem is confirmed on the basis of the explicit expressions of the transition amplitudes (see (\ref{eq:A-+}) and also refer to \cite{Suzuki2018}), which are derived from the analytic solution of the quantum two-level system \cite{MessinaNakazato2014}.
What is stressed here is that these expressions enable us to evaluate such transition amplitudes directly in any physical situation, from the adiabatic to diabatic cases.
For example, the deviation from the adiabatic limit is evaluated analytically and it is confirmed that the parameter characterizing the adiabaticity is given by the ratio between the fundamental frequency of the variation of the Hamiltonian and the frequency corresponding to its energy.
The observation is in accordance with the condition usually considered necessary for the adiabatic approximation to be valid $\hbar|_t\langle-|\dot H|+\rangle_t|/(E_-(t)-E_+(t))^2\ll1$.
Furthermore, we have made it clear that the necessary and sufficient condition that the instantaneous eigenstate of the Hamiltonian remains in the corresponding eigenstate at any time is that the speed of variation of the eigenstate is negligible compared with the frequency corresponding to the energy of the system. 
This condition is not the same as the traditional one just mentioned above and is free from the insufficiency found for the latter \cite{MarzlinSanders2004}. 
It is also remarked here that, since the conditions $\dot\theta\sim0,\,\dot\phi_\omega\sim0$ actually follow from $\dot\theta\sin\theta\sim0,\,\dot\phi_\omega\tan\theta_0\sim0$ under the assumption of nonvanishing angles  $\theta\not=0,\,\theta_0\not=0$, there is an exceptional case where $\theta\sim\theta_0$ is vanishingly small, for which no further conditions are required for other quantities \cite{Du2008}.
These observations are made possible owing to the explicit parametrization of transition amplitudes and further applications to other fundamental issues, including, e.g., the Landau--Zenner transition \cite{LZ}, will be published elsewhere.


\begin{thebibliography}{99}
\bibitem{BornFock1928}
M. Born and V. Fock, Z.\ Phys.\ {\bf51}, 165 (1928).  

\bibitem{Kato1950}
T. Kato, J. Phys.\ Soc.\ Jpn.\ {\bf5}, 435 (1950).

\bibitem{Messiah1962}
A. Messiah, {\it Quantum Mechanics\/}, (North-Holland Pub.\ Co., Amsterdam, 1962).

\bibitem{LZ}
L. D. Landau, Phys.\ Z. Sowjetunion {\bf2}, 46 (1932);
C. Zenner, Proc.\ R. Soc.\ A {\bf137}, 696 (1932).

\bibitem{GL}
M. Gell-Mann and F. Low, Phys.\ Rev.\ {\bf84}, 350 (1951).

\bibitem{Berry}
M. V. Berry, Proc.\ R. Soc.\ A {\bf392}, 45 (1984).

\bibitem{QC}
E. Farhi {\it et al\/}., Science {\bf291}, 472 (2001).

\bibitem{QZ}
P. Facchi and S. Pascazio, Phys.\ Rev.\ Lett.\ {\bf89}, 080401 (2002).

\bibitem{MarzlinSanders2004}
K. P. Marzlin and B. C. Sanders, Phys.\ Rev.\ Lett.\ {\bf93}, 160408 (2004);
D. M. Tong, K. Singh, L. C. Kwek, and C. H. Oh, {\it ibid\/}.\ {\bf95}, 110407 (2005).

\bibitem{Tong2007}
D. M. Tong, K. Singh, L. C. Kwek, and C. H. Oh, Phys.\ Rev.\ Lett.\ {\bf98}, 150402 (2007);
R. MacKenzie, A. Morin-Duchesne, H. Paquette, and J. Pinel, Phys.\ Rev.\ A {\bf76}, 044102 (2007); 
G. Rigolin, G. Ortiz, and V. H. Ponce, {\it ibid\/}.\ {\bf78}, 052508 (2008); 
M. H. S. Amin, Phys.\ Rev.\ Lett.\ {\bf102}, 220401 (2009);
D. Comparat, Phys.\ Rev.\ A {\bf80}, 012106 (2009);
D. M.  Tong, Phys.\ Rev.\ Lett.\ {\bf104}, 120401 (2010);
M.  Zhao and J. Wu, {\it ibid\/}.\ {\bf106}, 138901 (2011);
D. Comparat, {\it ibid\/}.\ {\bf106}, 138902 (2011);
D. M.  Tong, {\it ibid\/}.\ {\bf106}, 138903 (2011);
J. Ortigoso, Phys.\ Rev.\ A {\bf86}, 032121 (2012);
D. Li, Laser Phys.\ Lett.\ {\bf13}, 055203 (2016).

\bibitem{Du2008}
J. Du, L. Hu, Y. Wang, J. Wu, M. Zhao, and D. Suter, Phys.\ Rev.\ Lett.\ {\bf101}, 060403 (2008). 

\bibitem{WangPlenio2016}
Z.-Y.~Wang and M. B. Plenio, Phys.\ Rev.\ A {\bf93}, 052107 (2016).

\bibitem{BarnesDasSarma}
E. Barnes and S. Das Sarma, Phys.\ Rev.\ Lett.\ {\bf109}, 060401 (2012);
E. Barnes, Phys.\ Rev.\ A {\bf88}, 013818 (2013).

\bibitem{MessinaNakazato2014}
A. Messina and H. Nakazato, J.\ Phys.\ A: Math.\ Theor.\ {\bf47}, 445302 (2014).

\bibitem{Grimaudo2016}
R. Grimaudo, A. Messina, and H. Nakazato, Phys.\ Rev.\ A {\bf94}, 022108 (2016);
R. Grimaudo, A. Messina, P. A. Ivanov, and N. V. Vitanov, J. Phys.\ A {\bf50}, 175301 (2017);
L. A. Markovich, R. Grimaudo, A. Messina, and H. Nakazato, Ann.\ Phys.\ {\bf385}, 522 (2017);
R. Grimaudo, Yu.\ Belousov, H. Nakazato, and A. Messina, 
{\it ibid\/}.\ (to be published, https://doi.org/10.1016/j.aop.2018.03.012);
R. Grimaudo, A. S. M. de Castro, H. Nakazato, and A. Messina, 
arXiv:1803.02086.

\bibitem{t-dependence}
Here and in the following $t$-dependence will not be explicitly written, all quantities are understood to be time dependent unless otherwise stated explicitly and dot ($\dot{}$) stands for derivative w.r.t.\ time.

\bibitem{Suzuki2018}
T. Suzuki, Graduation thesis, Waseda University, 2018 (in Japanese).


\end{thebibliography}
\end{document}